\documentclass[prd,showpacs,preprint,nofootinbib]{revtex4-1}
\usepackage{amsmath,amssymb}
\usepackage[colorlinks=true,linkcolor=blue]{hyperref}
\usepackage{graphicx}
\usepackage{subfigure}

\usepackage{ulem}
\def\/{\over}

\newcommand{\bea}{\begin{eqnarray}}
\newcommand{\eea}{\end{eqnarray}}
\newcommand{\beq}{\begin{equation}}
\newcommand{\eeq}{\end{equation}}

\begin{document}

\title{High frequency background gravitational waves from spontaneous emission of gravitons by hydrogen and helium }

\author{Jiawei Hu$\footnote{jwhu@hunnu.edu.cn}$ and Hongwei Yu$\footnote{Corresponding author: hwyu@hunnu.edu.cn}$}
\affiliation{
$^1$ Department of Physics and Synergetic Innovation Center for Quantum Effects and Applications, Hunan Normal University, Changsha, Hunan 410081, China}

\begin{abstract}

A direct consequence of quantization of gravity would be the existence of gravitons. Therefore, spontaneous transition of an atom from an excited state to a lower-lying energy state accompanied with the emission of a graviton is expected. In this paper, we take the gravitons emitted by hydrogen and helium in the Universe after recombination as a possible source of high frequency background gravitational waves, and calculate the energy density spectrum. Explicit calculations show that the most prominent contribution comes from the $3d-1s$ transition of singly ionized helium $\mathrm{He}^{+}$, which gives a peak in frequency at  $\sim10^{13}$ Hz. Although the corresponding energy density is too small to be detected even with  state-of-the-art technology today,  we believe that the spontaneous emission of $\mathrm{He}^{+}$ is a natural source of high frequency gravitational waves, since it is a direct consequence if we accept that the basic quantum principles we are already familiar with apply as well to a quantum theory of gravity.

\end{abstract}

\maketitle

\section{Introduction}

Gravitational waves are ripples  of spacetime predicted in Einstein's general relativity  and they have been directly observed recently by LIGO \cite{ligo1,ligo2,ligo3,ligo4,ligo5,ligo6}. The detected signals were generated by binary black hole, or neutron star mergers, and the frequencies of these signals are in the regime of $10^1-10^2$ Hz. Although the frequencies of gravitational waves radiated by astrophysical sources are usually $\lesssim 10^{3}$ Hz \cite{low1,low2,low3}, there may exist other sources of gravitational waves whose frequencies are expected to be higher, such as thermal gravitational radiation from stars \cite{weinberg,Kogan},  astrophysical plasma interacting with electromagnetic radiation \cite{Servin}, primordial small mass black holes \cite{Kogan}, cosmic  strings \cite{cs2,cs3}, extra dimensions \cite{extra1,extra2,extra3}, string cosmology \cite{cs1,Gasperini03,Veneziano04}, inflation \cite{Peebles,Giovannini,Giovannini2}, preheating \cite{preheating,cai,fu}, cosmological phase transitions \cite{phase1,phase2}, and so on. However, many of the  possible sources  above are based on hypothetical theories which are yet to be verified.

In the present paper, we are concerned with another possible source of high frequency gravitational waves, i.e. the spontaneous emission of gravitons by hydrogen and helium after recombination. 
Recombination is a stage at which the free electrons became bound into hydrogen and helium
atoms, ending the scattering of photons. 
The atoms  get excited due to the background thermal radiation, and spontaneous emission occurs.
The emitted quanta can be photons, and can also be gravitons if gravity can be quantized. Similar to the electromagnetic spectrum emitted by atoms, the gravitational emission spectrum is expected to be a unique set of discrete spectral lines. However, due to the expansion of the Universe, the emitted gravitons are redshifted, so the spectrum expected to be observed today should be continuous instead of discrete.

In this paper, firstly, we will derive the graviton emission rate for hydrogen atoms. Let us note that this topic has been studied by several authors  \cite{weinberg,Lightman,Kiefer,Boughn}. In Weinberg's  book \cite{weinberg}, the result is obtained in a semi-classical approach by taking the quantum-mechanical  quadrupole transition matrix elements into the classical formula for the gravitational quadrupole  radiation. Following the same approach, Kiefer derived a result which is 4 orders of magnitude larger than Weinberg's result, and pointed out a numerical error in Weinberg's calculation \cite{Kiefer}. 
In Ref. \cite{Boughn}, the result is obtained based on the standard perturbation theory with the interaction Hamiltonian $H_I=-\frac{1}{2}h_{\mu\nu}T^{\mu\nu}$. For a hydrogen atom, this Hamiltonian describes the influence of the fluctuating gravitational fields on the electromagnetic interaction between the electron and the nucleus. Note that this Hamiltonian is not gauge invariant. Of course, a gauge dependent Hamiltonian does not necessarily  mean that computed physical observables depend on a particular gauge, but much care should be taken when a gauge-dependent quantity is involved in a  calculation to ensure that the outcome is not gauge dependent. 
In Ref. \cite{Boughn}, it has been shown that to ensure the gauge invariance of the result, one must consider how the electromagnetic stress-energy is changed in the presence of the gravitational field if one chooses to work in a frame which is not local inertial. Therefore, it is desirable to explore if we can address this issue with a gauge invariant Hamiltonian. This is what we are going to do in the present paper. 
We assume that the hydrogen atom is subjected to a bath of fluctuating quantum gravitational fields in vacuum which must exist if we accept that the basic quantum principles we are already familiar with apply as well to a quantum theory of gravity since they are  necessitated by the uncertainty principle.  Under the influence of quantum fluctuations of spacetime,  an instantaneous quadrupole moment will be induced in the hydrogen atom. 
We describe the interaction between the instantaneous quadrupole moment of a hydrogen atom  and the fluctuating gravitational fields with the quadrupolar interaction Hamiltonian \cite{Wu17}, which does not involve the electromagnetic degrees of freedom and is gauge invariant. So, here  not only  the issue is dealt  with a gauge invariant Hamiltonian, but also the underlying physical picture is different from that in Ref. \cite{Boughn}. 
We will work out the gravitational polarizability of the hydrogen atom and the emission rate based on the formalism first proposed by Dalibard, Dupont-Roc, and Cohen-Tannoudji (DDC) \cite{DDC1,DDC2}, which has also recently been applied to study the spontaneous excitation of an accelerated atom coupled with quantum fluctuations of spacetime \cite{Cheng19}. 
In the DDC approach, we can separately calculate the contributions of vacuum fluctuations and radiation reaction to the emission rate. 
Secondly, we will calculate the density spectrum of gravitational waves we observe today from the spontaneous emission of hydrogen atoms in the Universe. 
We also consider the contribution to the background gravitational waves from helium since it is the second most abundant element in the Universe. 
Natural units $\hbar=c=16\pi G=1$ will be used in this paper.

\section{Graviton emission rate for multilevel atoms}

We plan to study the  spontaneous emission of gravitons for a multilevel atom in interaction with a bath of fluctuating quantum gravitational fields in vacuum, 
and consider it as a source of high frequency gravitational waves. We assume that the 
 atoms comove with the expansion of the Universe, so the proper time of the atoms $\tau$ coincides with the cosmic time. Here, for simplicity, we neglect the effect of the cosmic expansion when calculating the  spontaneous emission rate, but take it into account later when calculating the density spectrum of gravitational waves. Therefore, we assume that the spacetime metric $g_{\mu\nu}$ can be expanded  as the  metric of the Minkowski spacetime $\eta_{\mu\nu}$ and a linearized perturbation $h_{\mu\nu}$. 
The Hamiltonian of a multilevel atom can be written as
\begin{eqnarray}
H_A(\tau)=\sum\limits_{n}\omega_n\sigma_{nn}(\tau),
\end{eqnarray}
where $\tau$ is the proper time, $\sigma_{nn}(\tau)=|n\rangle\langle n|$, and $|n\rangle$ denotes the eigenstate of the atom with energy $\omega_n$. 
The Hamiltonian of the quantum gravitational field takes the form
\begin{eqnarray}
H_F(\tau)=\sum\limits_{k}\omega_{\vec{k}}a^{\dagger}_{\vec{k}}a_{\vec{k}}\frac{dt}{d\tau},
\end{eqnarray}
in which $\vec{k}$ denotes the wave vector,  $a^{\dagger}_{\vec{k}}$ and $a_{\vec{k}}$ are the creation and annihilation operators respectively.  
The quadrupolar interaction between the  atom and the quantum gravitational fields can be expressed as
\begin{eqnarray}\label{eq3}
H_I(\tau)=-\frac{1}{2}Q_{ij}(\tau)E_{ij}(x(\tau)),
\end{eqnarray}
where 
\begin{equation}
Q_{ij}=\int d^3x\, \rho_M(x)\left(x_ix_j-\frac{1}{3}\delta_{ij}x_k x_k \right)\;
\end{equation}
is the gravitational quadrupole moment operator of the atom with $\rho_M(x)$ describing the mass distribution, and
$E_{ij}=C_{i0j0}$, with $C_{i0j0}$ being the Weyl tensor, which  can be regarded as the trace-free  part of the Riemann tensor $R_{i0j0}$. It has been shown in Ref. \cite{mtw} that the Riemann tensor is  gauge invariant  in linearized  theory of gravity, and so is  the Weyl tensor. Therefore, the quadrupolar interaction Hamiltonian (\ref{eq3}) we use here is  gauge invariant, which is different from the gauge dependent one used in Ref. \cite{Boughn}. The derivation of the interaction Hamiltonian can be found, e.g., in Ref. \cite{Wu17}.

The Heisenberg equations of motion for  dynamical variables of the atom and the gravitational field can be derived from the total Hamiltonian  $H=H_A(\tau)+H_F(\tau)+H_I(\tau)$. Following the DDC formalism \cite{DDC1,DDC2}, the equation of motion for the atomic energy $H_A$ can be separated into two parts, i.e. the vacuum fluctuations (VF) and the radiation reaction (RR) with the symmetric ordering of variables between the atom and field. Assume that initially the field is  in the vacuum state $|0\rangle$, and the atom is in state $|b\rangle$. The expectation of the rate of change of the atomic energy in state $|b\rangle$ can then be expressed as 
\begin{eqnarray}\label{evolution1}
\left\langle \frac{d}{d\tau}H_{A}(\tau)\right\rangle_{VF}&=&\frac{i}{2}\int_{\tau_0}^{\tau}d\tau'C_{ijkl}^F(x(\tau),x(\tau')) \frac{d}{d\tau}(\chi_{ijkl}^A)_b(\tau,\tau'),
\end{eqnarray}
\begin{eqnarray}\label{evolution2}
\left\langle \frac{d}{d\tau}H_{A}(\tau)\right\rangle_{RR}&=&\frac{i}{2}\int_{\tau_0}^{\tau}d\tau'\chi_{ijkl}^F(x(\tau),x(\tau')) \frac{d}{d\tau}(C_{ijkl}^A)_b(\tau,\tau'),
\end{eqnarray}
where
$|\rangle$=$|b,0\rangle$.
Here, $C_{ijkl}^F$ and $\chi_{ijkl}^F$ are the symmetric correlation function and linear susceptibility of the gravitational field, which are defined as
\begin{eqnarray}\label{fieldC}
&&C_{ijkl}^F(x(\tau),x(\tau'))=\frac{1}{2}\langle 0\left|\left\{E^F_{ij}(x(\tau)),E^F_{kl}(x(\tau'))\right\}\right|0\rangle,\\
&&\chi_{ijkl}^F(x(\tau),x(\tau'))=\frac{1}{2}\langle 0\left|\left[E^F_{ij}(x(\tau)),E^F_{kl}(x(\tau'))\right]\right|0\rangle.
\label{fieldX}
\end{eqnarray}
Similarly, $(C_{ijkl}^A)_b$ and $(\chi_{ijkl}^A)_b$  are the symmetric correlation function and the linear susceptibility of the atom, 
\begin{eqnarray}\label{atomc}
(C_{ijkl}^A)_b(\tau,\tau')&=&\frac{1}{2}\langle b\left|\left\{Q_{ij}^F(\tau),Q_{kl}^F(\tau')\right\}\right|b\rangle,\\
(\chi_{ijkl}^A)_b(\tau,\tau')&=&\frac{1}{2}\langle b\left|\left[Q_{ij}^F(\tau),Q_{kl}^F(\tau')\right]\right|b\rangle.\label{atomx}
\end{eqnarray}

We assume that the atom is static and located at the origin, so its trajectory can be  written as
\begin{eqnarray}
t(\tau)=\tau,\ \  x(\tau)=y(\tau)=z(\tau)=0,
\end{eqnarray}
where $\tau$ is the proper time. 
In the following, we work in the transverse-traceless (TT) gauge, so there are only spatial components $h_{ij}$ in the gravitational perturbations. In the quantum linearized theory of gravity, the quantized spacetime perturbation $h_{ij}$ can be written  as  \cite{Yu99}
\begin{equation}
h_{ij}=\int d^{3}\mathbf{k}\sum_{\lambda}
\frac{1}{2\omega(2\pi)^{3}}[a_{\mathbf{k},\lambda}e_{ij}(\mathbf{k},\lambda)e^{i(\mathbf{k}\cdot\mathbf{x}-\omega t)}+{\rm H.c.}],
\end{equation}
where  H.c. represents the Hermitian conjugate, $\lambda$ labels the polarization states, 
$e_{\mu\nu}(\mathbf{k},\lambda)$ is the  polarization tensor, and $\omega=|\mathbf{k}|=(k_{x}^{2}+k_{y}^{2}+k_{z}^{2})^{\frac{1}{2}}$. 
Direct calculations show that 
\begin{equation}\label{E}
E_{ij}=\frac{1}{2}\ddot{h}_{ij},
\end{equation}
where a dot denotes derivative with respect to $t$. The two-point function of $E_{ij}$  in the vacuum state can then be obtained as 
\begin{equation}\label{correlation1}
{\langle{0|E_{ij}(x)E_{kl}(x')|0}\rangle}=\frac{1}{8(2\pi)^{3}}\int d^{3}\mathbf{k}\sum_{\lambda}e_{ij}(\mathbf{k},\lambda)e_{kl}(\mathbf{k},\lambda)\,{\omega^{3}}e^{i\mathbf{k}\cdot(\mathbf{x}-\mathbf{x'})}e^{-i\omega(t-t')}.
\end{equation}
The summation of the polarization tensors in the transverse traceless gauge takes the following form \cite{Yu99}, 
\bea\label{correlation2}
\sum_{\lambda}e_{ij}(\mathbf{k},\lambda)e_{kl}(\mathbf{k},\lambda)&=&\delta_{ik}\delta_{jl}+\delta_{il}\delta_{jk}-\delta_{ij}\delta_{kl}+\hat{k_{i}}\hat{k_{j}}\hat{k_{k}}\hat{k_{l}}+\hat{k_{i}}\hat{k_{j}}\delta_{kl}+\hat{k_{k}}\hat{k_{l}}\delta_{ij}\nonumber\\
&&-\hat{k_{i}}\hat{k_{l}}\delta_{jk}-\hat{k_{i}}\hat{k_{k}}\delta_{jl}-\hat{k_{j}}\hat{k_{l}}\delta_{ik}-\hat{k_{j}}\hat{k_{k}}\delta_{il},
\eea
where $\hat{k_{i}}={k_{i}}/{k}$. 
According to Eqs. (\ref{fieldC}) and (\ref{fieldX}), the field statistical functions $C_{ijkl}^F$ and  $\chi_{ijkl}^F$  can be calculated as 
\begin{eqnarray}\nonumber
C_{1111}^F(x(\tau),x(\tau'))&=&-\frac{2}{\pi^2} \Delta^{+},\ \ \ \  \chi_{1111}^F(x(\tau),x(\tau'))=-\frac{2}{\pi^2} \Delta^{-},\\\nonumber
C_{1122}^F(x(\tau),x(\tau'))&=&\frac{1}{\pi^2}\Delta^{+},\ \ \ \ \ \  \chi_{1122}^F(x(\tau),x(\tau'))=\frac{1}{\pi^2}\Delta^{-},\\
C_{1212}^F(x(\tau),x(\tau'))&=&-\frac{3}{2\pi^2} \Delta^{+},  \ \ \chi_{1212}^F(x(\tau),x(\tau'))=-\frac{3}{2\pi^2} \Delta^{-},
\end{eqnarray}
where
\bea\label{sinh}
\Delta^{+}=\frac{1}{(\tau-\tau'-i\epsilon)^6}+\frac{1}{(\tau-\tau'+i\epsilon)^6},\qquad
\Delta^{-}=\frac{1}{(\tau-\tau'-i\epsilon)^6}-\frac{1}{(\tau-\tau'+i\epsilon)^6}.
\eea
The nonzero components of $C_{ijkl}^F$ satisfy the following relations,
\bea\nonumber\label{C-property}
C_{1111}^F&=&C_{2222}^F=C_{3333}^F,\\\nonumber 
C_{1122}^F&=&C_{2211}^F=C_{1133}^F=C_{3311}^F=C_{2233}^F=C_{3322}^F,\\ \nonumber
C_{1212}^F&=&C_{1221}^F=C_{2112}^F=C_{2121}^F=C_{1313}^F=C_{1331}^F\\
&=&C_{3113}^F=C_{3131}^F=C_{2323}^F=C_{2332}^F=C_{3223}^F=C_{3232}^F,
\eea
which are the same for $\chi_{ijkl}^F$. 
Inserting a complete set of states into Eqs. (\ref{atomc}) and (\ref{atomx}), it can be shown that the explicit forms of the statistical functions of the atom $(C_{ijkl}^A)_b(\tau,\tau')$  and $(\chi_{ijkl}^A)_b(\tau,\tau')$ are
\begin{eqnarray}
(C_{ijkl}^A)_b(\tau,\tau')&=&\frac{1}{2}\sum\limits_{\omega_{bd}}\biggl[\langle b|Q_{ij}^F(0)|d\rangle \langle d|Q_{kl}^F(0)|b\rangle e^{i\omega_{bd}(\tau-\tau')}\nonumber\\
&&\ \ \ \ \ \ +\langle b|Q_{kl}^F(0)|d\rangle \langle d|Q_{ij}^F(0)|b\rangle e^{-i\omega_{bd}(\tau-\tau')}\biggl],\\
(\chi_{ijkl}^A)_b(\tau,\tau')&=&
\frac{1}{2}\sum\limits_{\omega_{bd}}\biggl[\langle b|Q_{ij}^F(0)|d\rangle \langle d|Q_{kl}^F(0)|b\rangle e^{i\omega_{bd}(\tau-\tau')}\nonumber\\
&&\ \ \ \ \ \ -\langle b|Q_{kl}^F(0)|d\rangle \langle d|Q_{ij}^F(0)|b\rangle e^{-i\omega_{bd}(\tau-\tau')}\biggl],
\end{eqnarray}
respectively, where $\omega_{bd}=\omega_{b}-\omega_{d}$. 
With a substitution $u=\tau-\tau'$, and an extension of the range of integration to infinity\footnote{Here it is assumed that the time $\tau$ is much larger than the correlation time of the bath of fluctuating gravitational fields $\tau_c$, so the contribution to the integration comes mainly from the interval $[0,\tau_c]$, and it is safe to extend the integration to infinity \cite{DDC1,DDC2}.}, the contributions of vacuum fluctuations and radiation reaction to the average rate of change of the atomic energy can be calculated from Eqs. (\ref{evolution1}) and (\ref{evolution2}) as, 
\begin{eqnarray}\label{vf}
\left\langle \frac{d}{d\tau}H_{A}(\tau)\right\rangle_{VF}&=&-\frac{1}{4}\sum\limits_{\omega_{bd}}\omega_{bd}\biggl[|\langle b|Q_{11}^F(0)|d\rangle|^2+|\langle b|Q_{22}^F(0)|d\rangle|^2+|\langle b|Q_{33}^F(0)|d\rangle|^2\biggl] \mathcal{G}^F_{1111}\nonumber\\
&&-\frac{1}{4}\sum\limits_{\omega_{bd}}\omega_{bd}\biggl[\langle b|Q_{11}^F(0)|d\rangle \langle d|Q_{22}^F(0)|b\rangle+\langle b|Q_{22}^F(0)|d\rangle \langle d|Q_{11}^F(0)|b\rangle \nonumber\\ \nonumber
&&\qquad\quad\quad\ \ +\langle b|Q_{11}^F(0)|d\rangle \langle d|Q_{33}^F(0)|b\rangle+\langle b|Q_{33}^F(0)|d\rangle \langle d|Q_{11}^F(0)|b\rangle\\ \nonumber
&&\qquad\quad\quad\ \ +\langle b|Q_{22}^F(0)|d\rangle \langle d|Q_{33}^F(0)|b\rangle+\langle b|Q_{33}^F(0)|d\rangle \langle d|Q_{22}^F(0)|b\rangle\biggl] \mathcal{G}^F_{1122}\\ 
&&-\sum\limits_{\omega_{bd}}\omega_{bd}\biggl[|\langle b|Q_{12}^F(0)|d\rangle|^2+|\langle b|Q_{13}^F(0)|d\rangle|^2+|\langle b|Q_{23}^F(0)|d\rangle|^2\biggl]\mathcal{G}^F_{1212}
\end{eqnarray}
and
\begin{eqnarray}\label{rr}
\left\langle \frac{d}{d\tau}H_{A}(\tau)\right\rangle_{RR}&=&-\frac{1}{4}\sum\limits_{\omega_{bd}}\omega_{bd}\biggl[|\langle b|Q_{11}^F(0)|d\rangle|^2+|\langle b|Q_{22}^F(0)|d\rangle|^2+|\langle b|Q_{33}^F(0)|d\rangle|^2\biggl] \mathcal{K}^F_{1111}\nonumber\\ \nonumber
&&-\frac{1}{4}\sum\limits_{\omega_{bd}}\omega_{bd}\biggl[\langle b|Q_{11}^F(0)|d\rangle \langle d|Q_{22}^F(0)|b\rangle+\langle b|Q_{22}^F(0)|d\rangle \langle d|Q_{11}^F(0)|b\rangle\\ \nonumber
&&\qquad\quad\quad\ \ +\langle b|Q_{11}^F(0)|d\rangle \langle d|Q_{33}^F(0)|b\rangle+\langle b|Q_{33}^F(0)|d\rangle \langle d|Q_{11}^F(0)|b\rangle\\ \nonumber
&&\qquad\quad\quad\ \ +\langle b|Q_{22}^F(0)|d\rangle \langle d|Q_{33}^F(0)|b\rangle+\langle b|Q_{33}^F(0)|d\rangle \langle d|Q_{22}^F(0)|b\rangle\biggl] \mathcal{K}^F_{1122}\\ 
&&-\sum\limits_{\omega_{bd}}\omega_{bd}\biggl[|\langle b|Q_{12}^F(0)|d\rangle|^2+|\langle b|Q_{13}^F(0)|d\rangle|^2+|\langle b|Q_{23}^F(0)|d\rangle|^2\biggl] \mathcal{K}^F_{1212},
\end{eqnarray}
where
\begin{eqnarray}
\mathcal{G}^F_{ijkl}=\int_{-\infty}^{\infty}du\  e^{i\omega_{bd}u}C_{ijkl}^F(u) ,\ \ \ \mathcal{K}^F_{ijkl}=\int_{-\infty}^{\infty}du\  e^{i\omega_{bd}u}\chi_{ijkl}^F(u) 
\end{eqnarray}
are the Fourier transforms of $C_{ijkl}^F$ and $\chi_{ijkl}^F$. 
Note that the quadrupole-dependent terms in Eqs. (\ref{vf}) and (\ref{rr}) are equal. This 
leads to the cancellation of the contributions from vacuum fluctuations and radiation reaction for an inertial atom when $\omega_{bd}<0$, i.e. transitions to higher-lying levels are not allowed for inertial atoms in vacuum, as expected. When $\omega_{bd}>0$, the total average rate of change of the atomic energy is 
\begin{eqnarray}\label{tot}
\left\langle \frac{d}{d\tau}H_{A}(\tau)\right\rangle
&=&-\frac{2\hbar G}{15 c^5}\sum\limits_{\omega_{bd}>0}
\omega_{bd}^7\big(2\alpha_{1111}+2\alpha_{2222}+2\alpha_{3333}-\alpha_{1122}-\alpha_{2211}-\alpha_{1133}-\alpha_{3311} \nonumber\\
&& -\alpha_{2233}-\alpha_{3322}+6\alpha_{1212}+6\alpha_{1313}+6\alpha_{2323}\big).
\end{eqnarray}
Here we have returned to the SI units, and have defined the gravitational polarizability as $\alpha_{ijkl}=\langle b|Q_{ij}^F(0)|d\rangle \langle d|Q_{kl}^F(0)|b\rangle/\hbar\omega_{bd}$.

\section{The density spectrum of gravitational waves}

First, we investigate the density spectrum of background gravitational waves emitted by hydrogen atoms. 
We assume that the hydrogen atoms are in thermal equilibrium with the background  radiation and  satisfy the Boltzmann distribution, so the population decreases significantly as the  principal quantum number $n$ increases. On the other hand, gravitons are expected to be spin-2 particles, so the change of the orbital angular momentum quantum number $\Delta l$ should be 2 after  transition. Therefore, the most prominent process would be the transition from $3d$ to $1s$. 

The wavefunctions of hydrogen atoms in $3d$ and $1s$ states are 
\begin{equation}
\Psi_{1s}=\frac{1}{\sqrt{\pi}a^{3/2}}e^{-r/a},\qquad
\Psi_{3d}=\frac{1}{162\sqrt{\pi}a^{3/2}}\left(\frac{r^2}{a^2}\right)e^{-r/3a}
          \sin^2\theta \; e^{2i\phi},
\end{equation}
respectively. Direct calculations show that 
\begin{equation}\label{eq25}
\alpha_{1111}=\alpha_{2222}=\alpha_{1122}=\alpha_{2211}=\frac{243 }{8192} 
\frac{m_e a^2}{\omega_{31}^2},
\end{equation}
\begin{equation}
\alpha_{3333}=\frac{243 }{2048} \frac{m_e a^2}{\omega_{31}^2},
\end{equation}
\begin{equation}
\alpha_{1133}=\alpha_{3311}=\alpha_{2233}=\alpha_{3322}=-\frac{243 }{4096} 
\frac{m_e a^2}{\omega_{31}^2},
\end{equation}
\begin{equation}\label{eq28}
\alpha_{1212}=\alpha_{1313}=\alpha_{2323}=0,
\end{equation}
where $m_e$ is the mass of an electron, and $a$ is the Bohr radius. Therefore,
\begin{eqnarray}
\left\langle \frac{d}{d\tau}H_{A}(\tau)\right\rangle
&=&-\frac{3^8 G m_e^2 a^4 \omega_{bd}^6}{5\times2^{13}  c^5},
\end{eqnarray}
which agrees with the previous results derived from a semi-classical approach \cite{Kiefer}, and the standard perturbation theory \cite{Boughn}. 
At first glance, it may seem puzzling  that the rate of change of the atomic energy $\left\langle \frac{d}{d\tau}H_{A}(\tau)\right\rangle$ is proportional $a^4$ while the polarizability $\alpha_{ijkl}$ is proportional to $a^2$. 
Recall that the Bohr radius $a=\frac{4\pi\epsilon_0\hbar^2}{m_e e^2}$, and the transition frequency $\omega_{mn}=-\frac{m_e e^4}{2(4\pi\epsilon_0)^2\hbar^3}\left(\frac{1}{m^2}-\frac{1}{n^2}\right)$, so the Bohr radius $a$, the transition frequency $\omega_{bd}$, and the mass of electrons $m_e$ are not independent physical quantities. We write the rate of change of the atomic energy and the polarizability in the way above, since we are trying to avoid the Coulomb constant $\frac{1}{4\pi\epsilon_0}$ and the charge of electron $e$ in an  issue with gravitation as the main concern.

At a redshift $z\sim 1100$,  free electrons and protons became bound to form  hydrogen atoms,  and the Universe became transparent, which is known as recombination \cite{Cosmology}. In this paper, we consider gravitational waves produced by  the spontaneous emission of gravitons of hydrogen atoms in the Universe after recombination. Due to the expansion of the Universe, the gravitons emitted by hydrogen atoms are redshifted, and the redshifts are different for gravitons emitted at different time. Therefore, the spectrum expected to be observed today should be continuous instead of a series of discrete frequencies.

In the following, we calculate the  current energy density $\rho$ normalized with respect to the critical energy density $\rho_c=\frac{3H_0^2}{8\pi G}$, i.e. 
\begin{equation}\label{rho_c}
\Omega=\frac{1}{\rho_c}\frac{d\rho}{d\ln\omega}.
\end{equation}
The gravitational energy density emitted by hydrogen atoms from the time $\tau$ to $\tau+d\tau$  can be expressed as 
\begin{equation}\label{d_rho}
d\rho=-\frac{1}{(1+z)^4} N(z)\left\langle \frac{d H_{A}}{d\tau}\right\rangle  d\tau, 
\end{equation}
where $N(z)$ is the number density of hydrogen atoms in the $3d$ state. Here $\rho$ represents the  energy density at the current epoch $z=0$, which scales as $\frac{1}{(1+z)^4}$ due to the volume dilution as well as the redshift caused by the cosmic expansion. The minus sign indicates that  a decrease in the atomic energy means an increase in the gravitational wave energy. 
For simplicity, we assume that all atoms (ordinary matter) today are hydrogen atoms, and the number is conserved during the expansion of the Universe. Therefore, the number density of hydrogen atoms in the $3d$ state at redshift $z$ can be estimated as
\begin{equation}
N(z)\approx \frac{\rho_c P_{B}P_{3d}(z)}{m_H c^2} \left(1+z\right)^3,
\end{equation}
where $m_H$ is the mass of a hydrogen atom,  $P_B\approx 5\%$ is the current percentage of the baryonic matter, and $P_{3d}(z)$ is the percentage of hydrogen atoms in the $3d$ state. Here, we neglect the atoms  with principal quantum number $n \geq 5$ since the percentage is extremely small. The percentage of atoms in the $3d$ state can then be expressed as
\begin{equation}
P_{3d}(z) \approx \frac{n_{3d}(z)}{\sum_{n=1}^4\sum_{l=0}^{n-1}n_{ns}(z)}.
\end{equation}
Here 
\begin{equation}
n_{nl}(z)=(2l+1)n_{1s}(z) e^{-\frac{B_1-B_n}{k_B T(z)}},
\end{equation}
where  $B_n=13.6/n^2$ eV is the binding energy, $k_B$ is the Boltzmann constant, $n_{1s}$ is the population in the $1s$ state, and 
\begin{equation}\label{temp}
T(z)=(1+z) T_0\,,
\end{equation}
is the temperature  of the Universe when gravitons  are emitted at redshift $z$, with $T_0=2.73$ K  the temperature  of the Universe today. 
On the other hand, according to the definition of the Hubble constant $H$, it can be derived that 
\begin{equation}\label{36}
H d\tau=\frac{d\omega}{\omega}=d\ln\omega.
\end{equation}
Taking Eqs. (\ref{d_rho})-(\ref{36})  into Eq. (\ref{rho_c}), we have
\begin{equation}
\Omega=\frac{3^8\, G m_e^2 a^4 \omega_{31}^5 P_{B} P_{3d} \,\omega}{5\times2^{13}\, m_H c^7 H},
\end{equation}

\begin{figure}[htbp]
\centering
\subfigure{\includegraphics[width=0.49\textwidth]{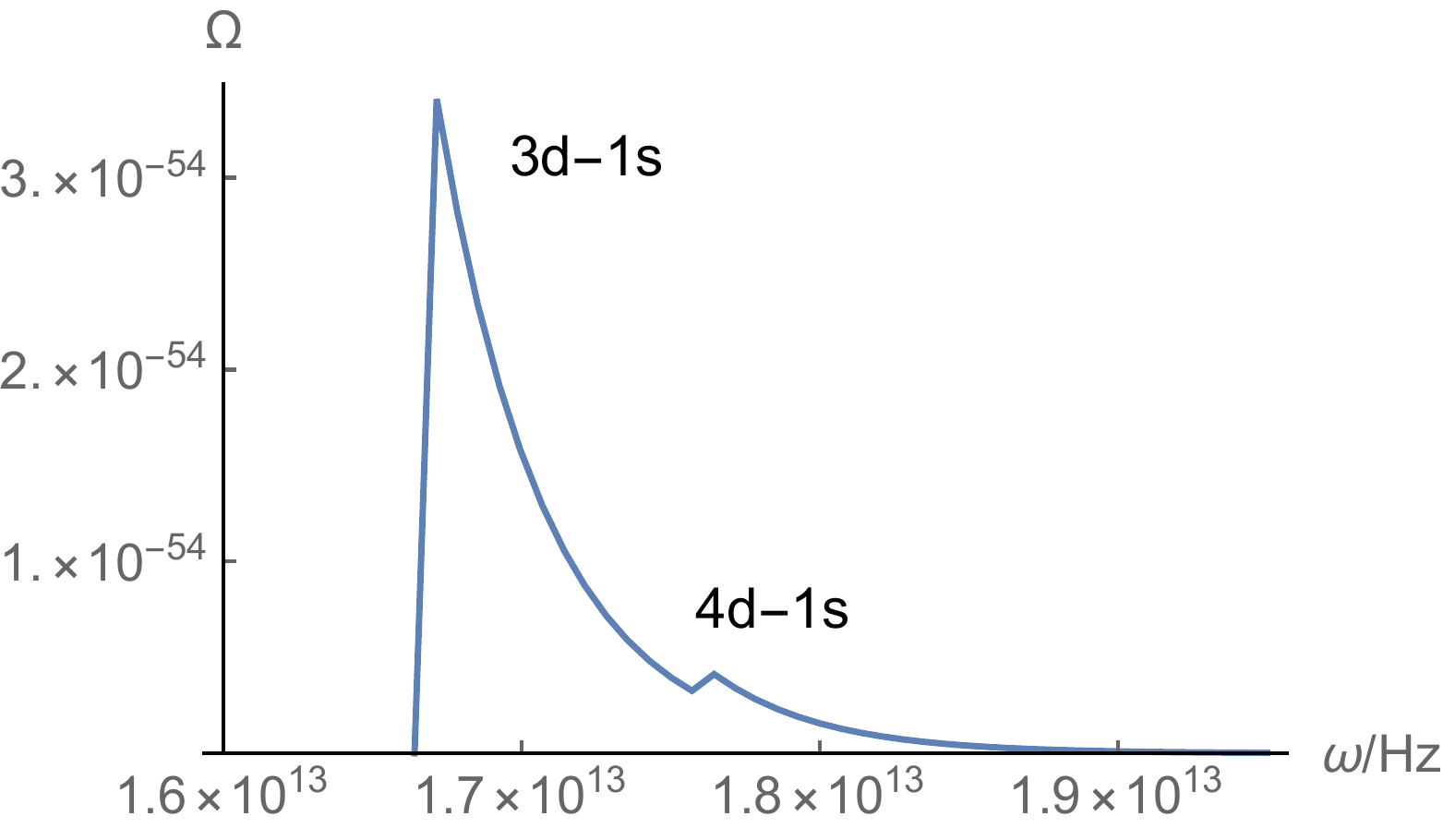}}
\subfigure{\includegraphics[width=0.49\textwidth]{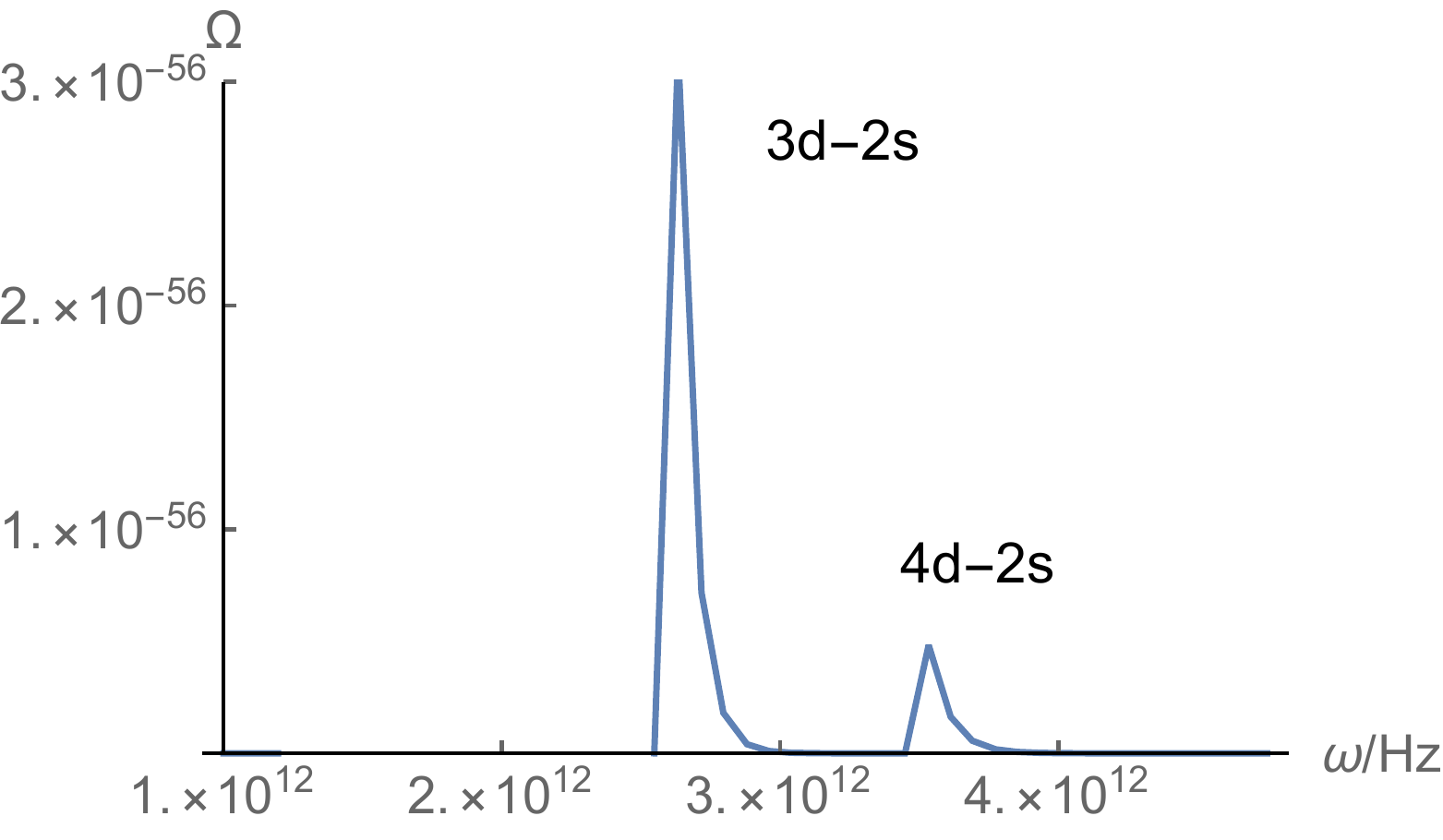}}
\caption{The energy density spectrum of gravitational waves from spontaneous emission of hydrogen atoms.}
\label{pic}
\end{figure}

Following the same procedures, we have considered all possible transitions for hydrogen atoms with the principal quantum number up to $n=4$. The results are shown in Fig. \ref{pic}. As expected, the dominant contribution comes from the $3d-1s$ transition at the redshift $z\sim 1100$, which corresponds to the peak in frequency at $\omega=1.67\times10^{13}$ Hz in Fig. \ref{pic} (left), and the relative energy density is $\sim 10^{-54}$. As $\omega$ increases, the energy density drops significantly. Physically, this means that the population of hydrogen atoms in the $3d$ state significantly decreases as the Universe cools down. There is also a small peak in frequency at $\omega=1.76\times10^{13}$ Hz in Fig. \ref{pic} (left), due to the $4d-1s$ transition, and the energy density is one order of magnitude smaller than the $3d-1s$ one, since the population of hydrogen atoms in the $4d$ state is smaller than that in the $3d$ state.  
In Fig. \ref{pic} (right), we show that in the $10^{12}$ Hz regime, there are two peaks corresponding to the $3d-2s$ and $4d-2s$ transitions, and the relative energy density is $\sim 10^{-56}$. Since the transition frequencies are smaller, the energy density is two orders of magnitude smaller than the  $3d-1s$ one. Other possible transitions include $4d-3s$, $4f-2p$ and $4f-3p$. Since the corresponding energy density are much smaller, we do not show here explicitly.

Apart from hydrogen, helium is the second most abundant element in the Universe,  constituting $\sim 24\%$  of the baryonic matter. The binding energy of helium is larger than that of hydrogen, so the recombination comes earlier. The recombination takes place in two steps. The recombination of singly ionized helium $\mathrm{He}^{+}$ takes place around redshift $z\approx 6000$, and the  recombination of neutral helium takes place around redshift $z\approx2000$ \cite{he1}. Here,  we consider only the contribution from $\mathrm{He}^{+}$ ions, since they recombine earlier and therefore the population of higher-lying excited states is larger, and they have a longer time to emit. 
Since a singly ionized helium $\mathrm{He}^{+}$ is a hydrogen-like atom, following the same calculations which have been done in the case of hydrogen atoms, we find a spectrum whose shape is similar to that of the hydrogen atoms but the signal is much stronger. Here, the dominant contribution comes from the $3d-1s$ transition of  $\mathrm{He}^{+}$ at the redshift $z\sim 6000$, which gives a peak in frequency at $\omega=1.22\times10^{13}$ Hz, and the relative energy density is $\sim 10^{-48}$, which is 6 orders of magnitude larger than that  from hydrogen atoms. 
This significant difference mainly comes from a much larger population of  higher-lying excited states since the recombination of $\mathrm{He}^{+}$ is much earlier and therefore the Universe is much hotter.

\section{Summary}

In summary, we take the gravitons emitted by hydrogen and helium in the Universe after recombination as a possible source of high frequency gravitational waves. In order to calculate the energy density spectrum, we first obtain the transition rate for multilevel atoms in interaction with  a bath of fluctuating quantum gravitational fields using the DDC formalism in the framework of the quantum linearized theory of gravity. Then we derive the energy density spectrum expected to be observed today.  Explicit calculations show that the most prominent contribution comes from the $3d-1s$ transition singly ionized helium $\mathrm{He}^{+}$, which gives a peak frequency  at  $\omega \sim 10^{13}$ Hz. Since the population  in excited states decreases significantly as the temperature of the Universe cools down, the energy density quickly decreases as the frequency we observe today increases. Although  far from the  precision of measurement today, we believe that  the spontaneous emission of $\mathrm{He}^{+}$  is a natural source of high frequency gravitational waves, since it is a direct consequence if we accept that the basic quantum principles we are already familiar with apply as well to a quantum theory of gravity and no hypothetical theories are involved.   

\begin{acknowledgments}

We thank the anonymous referees very much for their valuable comments and suggestions. 
This work was supported in part by the NSFC under Grants No. 11805063, No. 11690034, and No. 12075084, and the Hunan Provincial Natural Science Foundation of China under Grant No. 2020JJ3026. 

\end{acknowledgments}

\end{document}